# Focus Correction in MR thermography for Precise Targeting in Focused Ultrasound Thalamotomy for Essential Tremor

Authors:

[1,2]Chang-Sheng Mei, [3]Shenyan Zong, [2]Bruno Madore, [2]Garth R. Cosgrove, [2]Nathan J. McDannold

Affiliations:

[1]Department of Physics, Soochow University, Taipei

[2]Department of Radiology, Harvard Medical School, Brigham and Women's Hospital, Boston, MA, USA

[3]School of Biomedical Engineering, Shanghai Jiao Tong University, Shanghai

Correspondence to:

Chang-Sheng Mei

meic@bwh.harvard.edu

Word Count:   2501 (excluding abstract, reference and captions)

ABSTRACT

**Purpose:** Thermal ablation by focused ultrasound (FUS) is an invasive surgery for essential tremor (ET) treatment. With the guidance of magnetic resonance (MR) thermometry, the sonicated thalamic target on focus can be successfully monitored. However, the spatial discrepancy induced by temperature changes between the target coordinate and the hotspot as seen in MR thermography often occurs. The goal of this work is to adjust the location of ablation focus to achieve the accurate targeting in ET treatments.

**Methods:** Two causes, chemical shift and k-space shift, are accounted for the observed spatial misregistration. The well-versed chemical shift-caused displacement results from the fact that the heating can spatially differ the resonance frequency. Such spatial shift was corrected pixel by pixel using the obtained field map around focus. The less-known temperature errors shifting the focus location come from the hotspot gradient changes. The temperature-induced TE variation map was calculated to compensate the center location offset of focus. The proposed two-step correction procedures were initially validated in heating simulation and FUS phantom heating experiments. From the clinical data, 121 sonication locations from 7 ET patients were also analyzed and compared for evaluating target location accuracy after correction.

**Results:** In phantom experiments, a total of approximate 1mm shift caused by chemical shift and k-space center offset was adjusted using the field map and TE error map. In analysis of 121 ET treatments, a close linearity coefficient of $R^2=0.852$ was found between temperature increase and focus shift. From the fitted slope, an error of about 0.5mm in focus location was expected for every 10°C of temperature elevation. In comparison of Bland-Altman plots, the mean temperature error was reduced to -0.05°C, as opposed to that error of -0.11°C obtained using only a chemical shift correction.

**Conclusion:** The spatial shifts of hotspot caused by field gradients and k-space shifts in temperature changes need be corrected in FUS thalamotomy treatment of ET symptoms. The proposed correction method can ensure that temperature hotspots as visualized on MR thermography match as closely as possible in location with lesions.

INTRODUCTION

Focused ultrasound (FUS) thalamotomy is an effective alternative to surgical resection for those drug-refractory essential tremors (ET) or Parkinson's tremors (PT). Magnetic resonance imaging (MRI) guidance provides the noninvasive manner to visualize the transcranial ablation treatments through image-based temperature measurements on target. In clinical, the generally accepted MR thermometry is to employ the proton resonance frequency (PRF) method. During the current surgery procedures, a series of low acoustic heating trials are progressively performed to adjust focus location. After that, the hotspot with a mass of heat deposition on target is generated to balance the excessive neural conduction in several acoustic deliveries, for the tremor relief on hands. As such, the accurate focusing of FUS beams is critical to enhance the thermal therapy and decrease the surgery duration, and the small size of targeted structure also requires the FUS focus accurately aims at the target. However, the intrinsic focus shift on spatial caused by temperature changes often occurs between the target coordinate and the hotspot observed in MR thermograph, as a potential source of safety concerns. During the current ET thalamotomy, a shifted distance of about 3mm frequently occurred, as seen in Fig. 1.

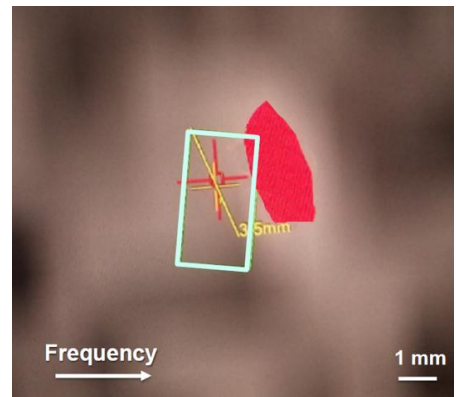

Fig. 1: A focus location offset case during the ET thalamotomy. A shift distance of about 3.5 mm was observed along the frequency encoding direction.

In PRF, the temperature-caused resonance frequency shift is utilized to obtain the temperature changes in thermal ablations. Under the magnetic field of 3 Tesla, the temperature change of 1°C on aqueous tissues can result in the resonance frequency offset of 1.28 Hz. And the higher temperature increase on a voxel will produce the wider frequency shift, thus making this voxel migrate to its adjacent location. Such resonance frequency offset is hazardous in PRF-based MR thermometry, because the low receiving bandwidth is usually applied in the temperature measurements for the improvement of temperature to noise ratio (TNR). Theoretically, the resonance frequency discrepancy by temperature belongs the chemical shift. The resulting artifacts generally manifest as the misalignment on image. In the view of MR thermography, the congener artifact is revealed as the hotspot displacement from the expected target. The chemical shift associated with heating is the well-known reason responsible for the focus spatial errors.

The rarely known reason that can account for the hotspot shift on spatial is the echo time variations pixel by pixel. PRF method often converts phase changes on MR images to temperature maps. The focus spot viewed from two-dimensional temperature maps always presents as the ellipse in case the imaged plane is paralleled to the ultrasound beams and circle while the monitored slice is perpendicular to the ultrasound beams. Both temperature distributions appear as the highest temperature at the center and low temperature on the edge. Therefore, the temperatures on hotspot change from zero to the hottest, and then descend to zero along the frequency encoding direction. With regards to its MR phase maps, the phase gradients are bipolar on the two sides of focus profile. As demonstrated on our previous work, the heating can lead to the TE errors around the focus region due to the spatial phase variations, and that the time of TE plays a significant role in converting the phase changes to temperature map. In result, the temperature errors on focus respectively exhibit as overestimated and underestimated on the two sides, leading to the spatial displacement of the focus. The infrequent reason for the focus shift is called the k-space offset.

The goal of this work was to compensate and correct the focus shift caused by the chemical shift and the k-space offset. In the first-step scenario, the temperature increases were reversely derived to the resonance frequency change maps using the imaging bandwidth, thus knowing the spatial pixel shift scope. The shifted distance on each pixel was used to move the focus back. In the second-step correction, the temperature map was translated into the TE error map through the phase gradient changes. The obtained pixel-to-pixel TE errors compensated the user-enter nominal TE value. Afterwards, the temperature-induced phase changes were re-convert to the temperature map, to get the more accurate hotspot location. The proposed correction method here was verified on the ET patients received the FUS thalamotomy, and the results demonstrated the focus accuracy can be improved after the chemical shift and k-space offset corrections.

THEORY

PRF thermometry, Chemical Shift and TE Errors

For a regular gradient echo (GRE) sequence, PRF-based thermometry uses the temperature-caused phase change on spatial - $\Delta\emptyset(\vec{r})$, to divide a factor involving with the echo time – TE, the magnetic field strength – B0, the gyromagnetic ratio for hydrogen – γ, and the PRF change coefficient – α. The PRF calculation equation was express as:

$$\Delta T(\vec{r}) = \frac{\Delta\emptyset(\vec{r})}{2\pi \cdot \gamma \cdot \alpha \cdot B0 \cdot TE} \quad [1]$$

where the $\Delta T(\vec{r})$ represents the spatial temperature change, and the constant coefficient, $\gamma$ and $\alpha$, are equal to 42.576 MHz/T and -0.01 ppm/°C, respectively. The TE is typically defined as the time interval from the excitation radio frequency (RF) pulse to the echo peak.

After calculation by Eq. [1], a temperature map without hotspot location correction can be obtained. The increased temperature $\Delta T(\vec{r})$ on a given pixel corresponds to the fixed resonance frequency shift, indicating the pixel location motion along the frequency encoding direction. The following Eq. [2] can obtain the shift quantity of the pixel on focus combining with the receiving bandwidth, BW:

$$\Delta x(\vec{r}) = \frac{\gamma \alpha B0 \cdot \Delta T(\vec{r})}{BW} \quad [2]$$

In Eq. [2], the $\Delta x(\vec{r})$ presents the pixel offset distance resulted from the heating along the frequency encoding direction. Other parameters are identical to those written in Eq. [1]. This is the so-call chemical shift-associated focus displacement.

Besides, the phase gradient on image domain can displace the "true" k-space center, to vary the spatial TE values around the hotspot region. Such errors on TE can be calculated from the phase gradient changes as follow:

$$\Delta TE(\vec{r}) = N_f \times \frac{\nabla_x(\Delta \emptyset(\vec{r}))}{2\pi \cdot BW} \quad [3]$$

where the $\Delta TE(\vec{r})$ stands for the spatial TE errors, the $\Delta \emptyset(\vec{r})$ indicates the phase changes before and after heating, and then the $\nabla_x(\Delta \emptyset(\vec{r}))$ refers to the phase gradients on phase difference map along the frequency encoding direction. The $N_f$ represents the frequency encoding number in the imaged field of view (FOV), and the BW also implies the receiving bandwidth. In doing so, the user-input TE in Eq. [1] can be corrected through Eq. [4]:

$$\widehat{TE}(\vec{r}) = TE + \Delta TE(\vec{r}) \quad [4]$$

The corrected echo time $\Delta TE(\vec{r})$ in Eq. [4] was used to replace the TE coefficient in Eq. [1], to compensate the shifted hotspot location again.

Correction Workflow

The diagram in Fig. 2 illustrates the steps involved in correcting these problems. The temperature map around the focus (Fig. 2a,b) is first corrected pixel-to-pixel using a spatial-shift map (Fig. 2c), based on the heating-induced chemical shift. The resulting map (Fig. 2d) is then corrected using a TE-error map (Fig. 2e), based on the pixel-by-pixel shift in k-space. Because the TE error on either side of the focus has opposite polarity (shown as red and blue in Fig. 2e), the temperature on one side of the focus is elevated while temperature on the other side is suppressed, effectively shifting the apparent location of the focus (Fig. 2f).

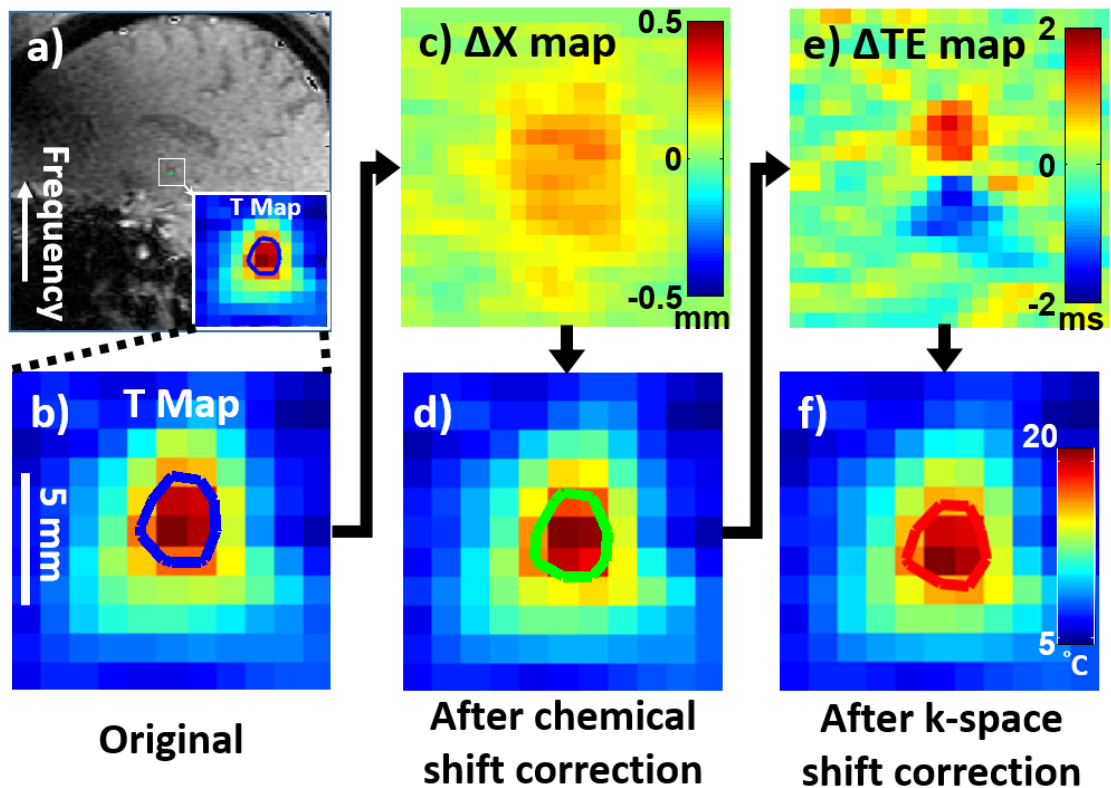

Fig. 2: Diagram illustrating the method to correct the heating-induced spatial error of focus in FUS thalamotomy4. An ET patient was treated with FUS in thalamus (a). The temperature map around focus (b) was first corrected based on a ΔX map (c), calculated from the heating-induced chemical shift. The resulting temperature map (d) was then corrected with a ΔTE map (e) calculated from heating-induced shifts in k-space5. A shift in the frequency-encoding direction can be observed between the uncorrected (blue contour) and fully corrected (red contour) heating; the shift of the centroid (15.9 °C) was calculated at 0.95mm.

Simulations

Having known the phase gradient causes temperature error, let's try to understand why the error in temperature would lead to spatial error in focus. The discussion below will focus on correcting the error in frequency encoding direction, since such error is bound to appear in the frequency encoding direction, as being described in the previous sessions. Unlike a global heating, such as submerging the phantom into a warm water, the focal heating from FUS is local and relatively small. Therefore, the temperature profile of focus is a Gaussian function, starting the temperature elevation from zero to the maximum and then back to zero in spatial domain. Therefore, the phase gradient caused by such heating profile is positive on one side and negative on the other side along frequency encoding axis. Accordingly, the temperature error caused by such gradient has different polarity on either side of the focus, i.e. temperature being underestimated on one side and overestimated on the other. After proper correction, the temperature on one side of the focus gets risen and the temperature on the other side gets reduced. Result of this is a spatial shift of the focus.

The spatial shift of focus caused by different polarity of temperature error is demonstrated in Fig. 3. In a Gaussian-simulated FUS heating, the temperature map with maximum heating of 15 °C is shown in Fig. 3**Error! Reference source not found.**a. The heating profile along presumable frequency encoding direction (horizontal) is inset on the upper left corner. On the left side, the heating causes positive phase gradient, while on the right side, the heating causes negative phase gradient. Consequently, the temperature is overestimated on the left side and underestimated on the right, giving a temperature error map with spatially different polarity on either side of focus, as shown in Fig. 3b. This is why the error of temperature causes spatial shift of focus. Fig. 3c shows the spatial shift from the original (blue) to the adjusted (red) curves. With overestimation on the left side of focus, the 'real' temperature on that side gets decreased. With underestimation on the right, the 'real' temperature on the right gets increased. As a result, the focus shifts a bit to the right.

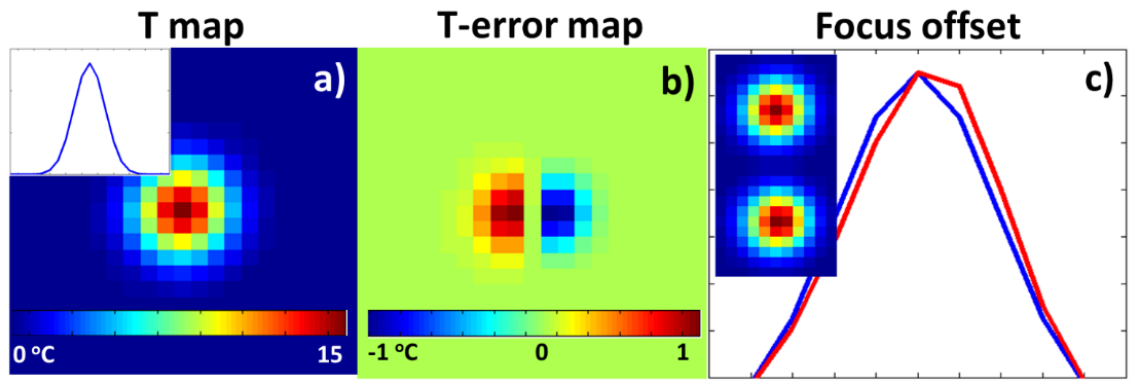

Fig. 3: Demonstration of how temperature error of focus would lead to focus offset. a) A temperature map of a FUS heating, simulated with Gaussian filter. Inset is the heating profile along horizontal direction. b) A temperature error map caused by the focal heating shows different polarity on either side of focus, leading to temperature overestimate on the left side and underestimate on the right. c) The temperature profile through hottest pixel is plotted, and a shift in horizontal direction can be observed between the original (blue) and the adjusted (red) curves. Inset shows the original T map (up) and the adjusted T map.

MATERIALS and METHODS

Phantom Experiments

Six FUS heating experiments were performed in a gel phantom (InSightec, Israel) without a skull in place using a hemispheric, 650-kHz, 1024-element, phased-array transducer (ExAblate Neuro, InSightec, Israel) while the temperature was monitored with a 3T MRI (Architec, GE, Milwaukee, WI). Imaging parameters are FOV = 28×28 cm2, TE/TR = 13.1/26.2 ms, matrix size = 128×256,

BW = ±5.68 kHz. Acoustic powers of 6, 12, and 30 W were delivered for 30 s for temperature elevations of 7, 12, and 22°C, respectively. In each power setting, two heating experiments were performed with the phase encoding direction reversed. The temperature maps with max heating less than 2°C were eliminated, leaving a total of 56 time frames for data analysis. Raw data from GE's "ScanArchive" were retrieved and analyzed with Matlab. In addition to the spatial correction due to chemical shift, the spatial errors caused by the spatial heating gradient (and therefore magnetic field gradient) were corrected in k-space, using a method previously reported.

Treatments of ET patients

A total of 7 patients received the transcranial FUS ablation treatments with 121 sonications, following the informed consent. The sonication was achieved using a hemispheric, 1024-elements, phase-array transducer working at 650 kHz (ExAblate Neuro 4000, InSightec, Israel) under the temperature monitor using a 3.0 T MR system (GE, Milwaukee, WI). The spoiled gradient echo (SPGR) sequence was used to perform MR thermometry with these settings: TR/TE = 27.8/12.9 ms, BW = ±5.68 kHz, FOV = 28×28 cm$^2$, matrix size = 128×256 with zero-filling to 256×256. The patients' sonication statistic and demographics were listed in Table 1. All 22 sonications were excluded following the rules: the too low temperature was produced (<5°C), or cavitation was detected, or the surgeon stopped the sonications out of safety concerns, or the patient himself did so. In this study, a total of 99 sonications were left for the further analysis. The data analysis included the calculation of the size of the focus shift for all sonications and perform a linear regression between the temperature increase and the shifted distance. The Bland-Altman plots were used to depict the focus profiles on temperature maps before and after correction.

| Patient demographics and sonication statistics | |
|---|---|
| | Study population |
| **Total number of patients** | 7 |
| **Total number of sonications** | 121 |
| **Gender** | |
| Male | 7 |
| Femal | 0 |
| **Age (y)** | |
| Mean ± SD | 67.9 ± 6.3 |
| Median (min, max) | 68 (58, 78) |
| **Sonication number per patient** | |
| Max T* < 5 °C | 3 ± 4 |
| 5 < Max T < 10 °C | 4 ± 2 |
| 10 < Max T < 15 °C | 4 ± 2 |
| Max T > 15 °C | 6 ± 3 |
| Total number | 17 ± 5 |
| **Focus shift (mm)** | |
| Max T < 5 °C | 0.6 ± 0.4** |
| 5 < Max T < 10 °C | 0.3 ± 0.1 |
| 10 < Max T < 15 °C | 0.5 ± 0.2 |
| Max T > 15 °C | 0.9 ± 0.2 |

*Max T is the maximum temperature in focus during sonication.
**Focus shift may not be correct if Max T < 5 °C (see context).

Table 1: Patient demographics and sonication statistics.

RESULTS

Phantom Experiments

Fig. 5 shows a strong linear correlation between temperature elevation and focus shift ($R^2=0.852$). 1.2mm of spatial error occurred at 22.7°C. For every 10°C of temperature elevation, the fitted slope predicts a focus error of about 0.5mm, which is consistent with clinical observations [2]. Images before and after correction are shown in Fig. 4. Fig. 4a shows the 2D heating focus from a FUS heating experiment. The 1D focal profile (red box in Fig. 4a) versus time is presented in Fig. 4b. We can visualize how the focus shifts upward with increasing temperature along the frequency encoding direction. After k-space shift correction, the focus remains symmetric for the 2D image (Fig. 4c) and for all time frames (Fig. 4d).

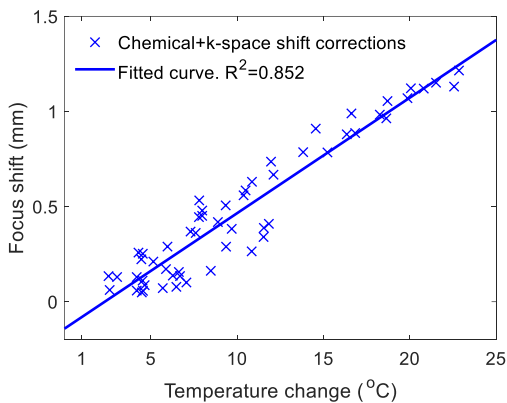

Fig. 5: There is a strong linear relationship (R2=0.852) between focus spatial errors and temperature rise. Focus offsets ~0.5 mm for every 10oC elevation.

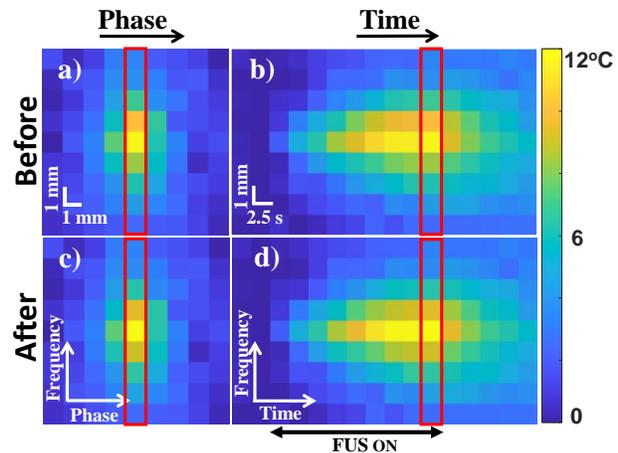

Fig. 4: The focus shifts upward along y direction (frequency encoding) with rising temperature in a) x-y b) y-t maps. The offsets are corrected with the proposed method as the profiles become symmetric in c),d).

Treatments of ET patients

Results for chemical shift and k-space shift correction are shown in Fig. 7 and Fig. 6. In Fig. 7a, the Δx map shows the spatial shift for pixels in the T map, due to a 'chemical shift' effect. In Fig. 7b, the TE error map can be used to correct temperature values: Note the different polarity on either side of the profile, causing overestimation on one side and underestimation on the other. The temperature elevation map T around the hot spot is shown in Fig. 6 without correction (Fig. 6a), using only the Δx correction from Fig. 7a (Fig. 6b, centroid shifts by 0.56 mm), or using both corrections from Fig. 7 (Fig. 6c, 0.56 + 0.39 mm = 0.95 mm shift). The resulting spatial shifts can also be visualized in Fig. 6d.

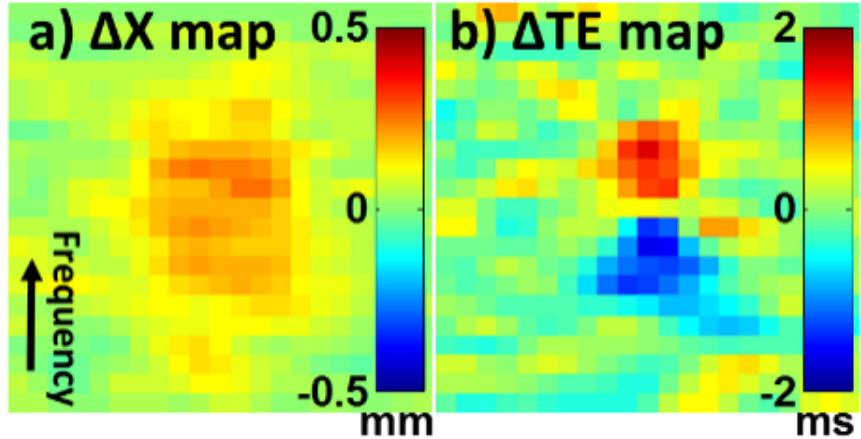

Fig. 7: Maps of spatial shifts in frequency direction (Δx) and TE error (ΔTE). Δx results from temperature-induced 'chemical shift' corrections, while ΔTE results from k-space shifts. Note the two lobes of opposite polarity in the ΔTE map, which caused temperature overestimation on one side of the profile and underestimation on the other side, combining into a net spatial shift in hot spot position.

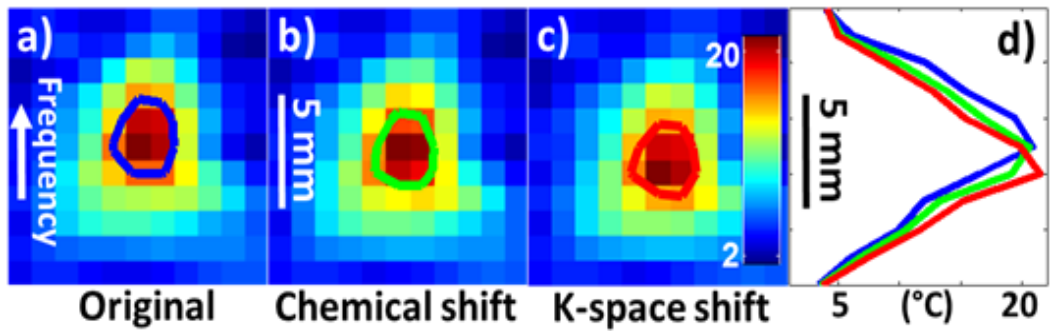

Fig. 6: Spatial shifts in the frequency encoding direction are shown and corrected. a) Original T map, b) T after chemical shift correction, and c) T map after both chemical shift and k-space shift correction. d) The temperature profile through hottest pixel is plotted, and a shift in the frequency-encoding direction can be observed between the uncorrected (blue) and fully corrected (red) curves, a shift of the centroid (15.9 °C) by 0.95 mm here.

A close linear correlation (R2=0.852) was found between temperature elevation and focus shift (blue line in Fig. 8). The maximum spatial error was 1.5mm and occurred at 28.4°C. From the fitted slope, an error of about 0.5mm in focus location is expected for every 10ºC of temperature elevation. As shown with red circles and line in Fig. 8, the chemical shift artifact could account for only about half of the observed shift, and effects from less-recognized k-space shifts must be included to account for the full observed error in focus location. A 5x5 ROI surrounding the focus, before and after the correction, wwa compared with scatterplots (Fig. 9a,b) and Bland-Altman plots (Fig. 9c,d). Looking at the Bland-Altman plots in Fig. 9c and 9d, including the k-space shift correction, as opposed to using only a chemical-shift correction, helped reduce the mean error from -0.11°C down to -0.05°C over the ROIs.

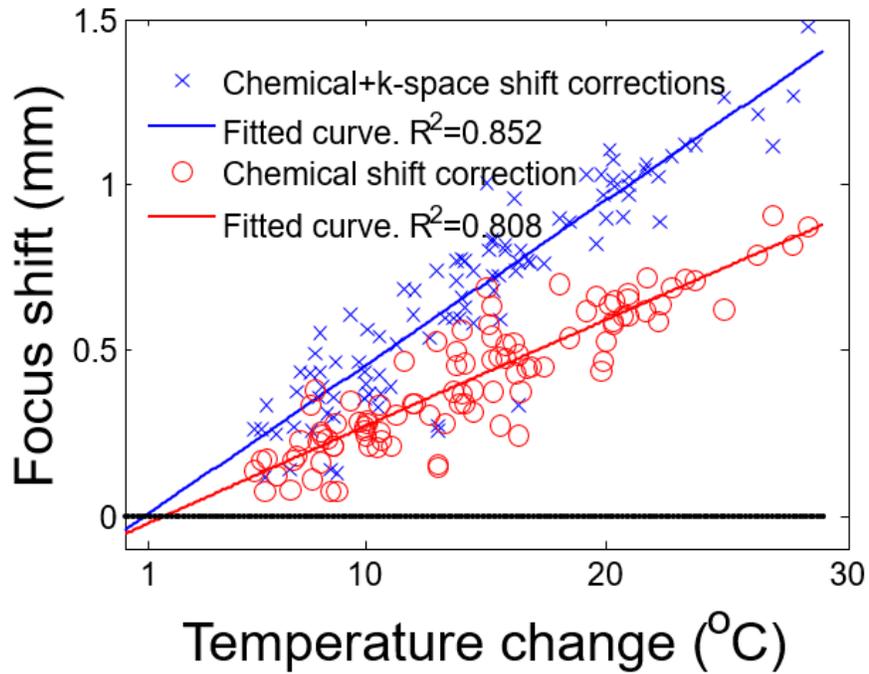

Fig. 8: Correlation between the maximum temperature changes at the focus acquired with MR thermometry and focus shifts computed with two correction methods from 99 sonications in 7 patients. Red circles represent the focus shift after chemical-shift correction, while the blue crosses represent the one after both corrections. Good linearity was observed for both scenarios, but the focus shifts with chemical-shift only did not reflect the shift as observed in the clinical setting, unlike the ~1mm shift handled by the proposed method when both types of correction were included. From the blue fitted line, focus was shifted by about 0.05 mm/°C.

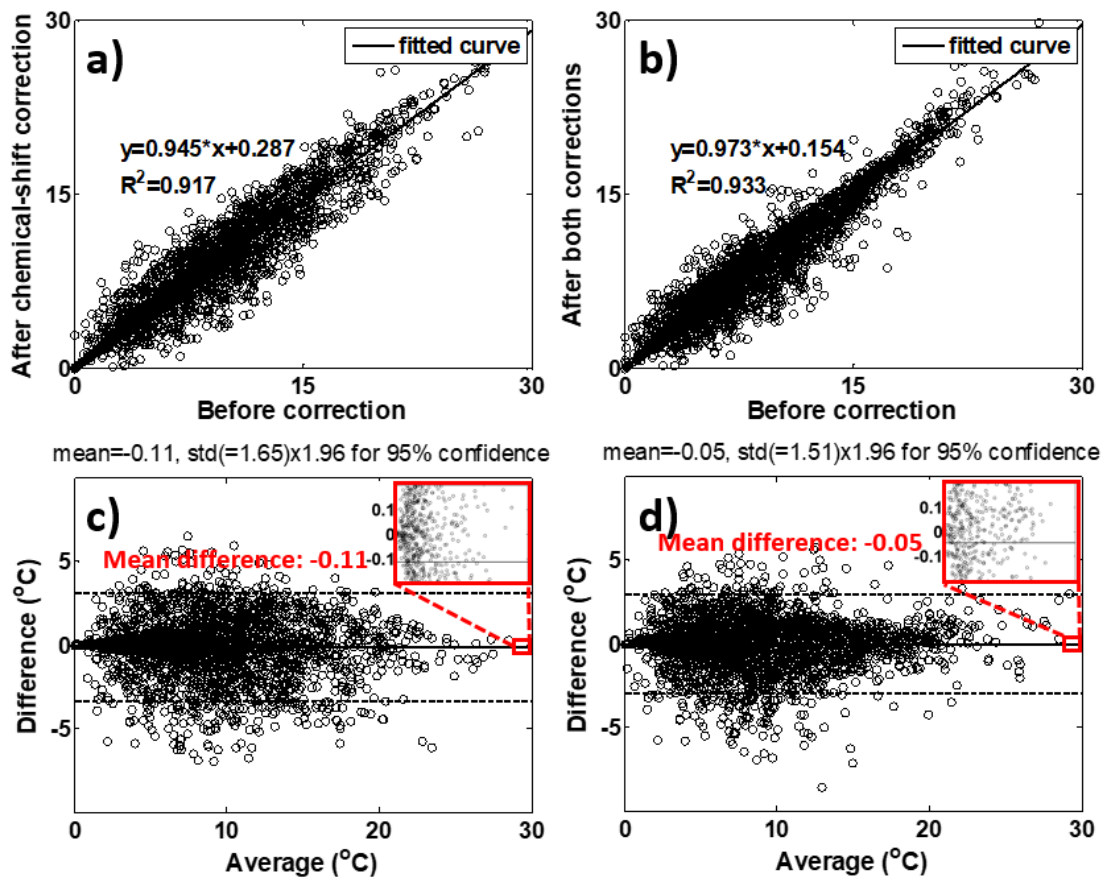

Fig. 9: Scatterplots (a, b) and Bland-Altman plots (c, d) comparing temperature measurements before and after correction are shown, for a 5x5 ROI centered at the focus, for 99 sonications involving 7 patients. a,c) Comparisons using chemical-shift correction only, and b,d) comparisons using both chemical-shift and k-space shift corrections. In a) and b) solid lines represent linear regression, while in c) and d) they represent the mean difference. Dashed lines mark the limits of the 95% confidence intervals. The mean difference between measurements for the chemical-shift correction scenario was -0.11°C, while it was reduced to -0.05°C when both corrections were applied.

DISCUSSION

This study investigated the impact of focus correction on the accuracy of MR thermometry during focused ultrasound (FUS) thalamotomy. Our findings demonstrate that correcting for spatial errors in the location of the thermal hotspots, as visualized with MRI, is crucial for precise targeting in FUS thalamotomy.

We observed a strong correlation between temperature elevation and focus shifts, with a 0.5mm shift for every 10°C increase. This finding underscores the significant impact of temperature changes on the accuracy of FUS targeting. Importantly, our analysis revealed that k-space shifts, induced by field gradients, contribute substantially to focus location errors. These effects, often overlooked in favor of the more recognized chemical shift artifacts, accounted for approximately half of the observed errors.

In Fig. 10a, T maps were overlaid onto the corresponding anatomical image, using red and green overlays for results without and with correction, respectively. As seen in Fig. 10b, spatial shifts did not appreciably affect temporal dynamics.

By accounting for both k-space shifts and chemical shift effects in our correction algorithm, we were able to significantly improve the accuracy of thermal hotspot localization. This improvement ensures a closer match between the visualized hotspots on MRI and the actual lesion location in the thalamus. Consequently, incorporating focus correction based on both k-space and chemical shift effects can enhance the precision and safety of FUS thalamotomy procedures.

Interestingly, while spatial shifts were effectively corrected, the temporal dynamics of the temperature maps remained unaffected. This observation suggests that our correction method specifically addresses spatial accuracy without altering the temporal characteristics of the thermal data.

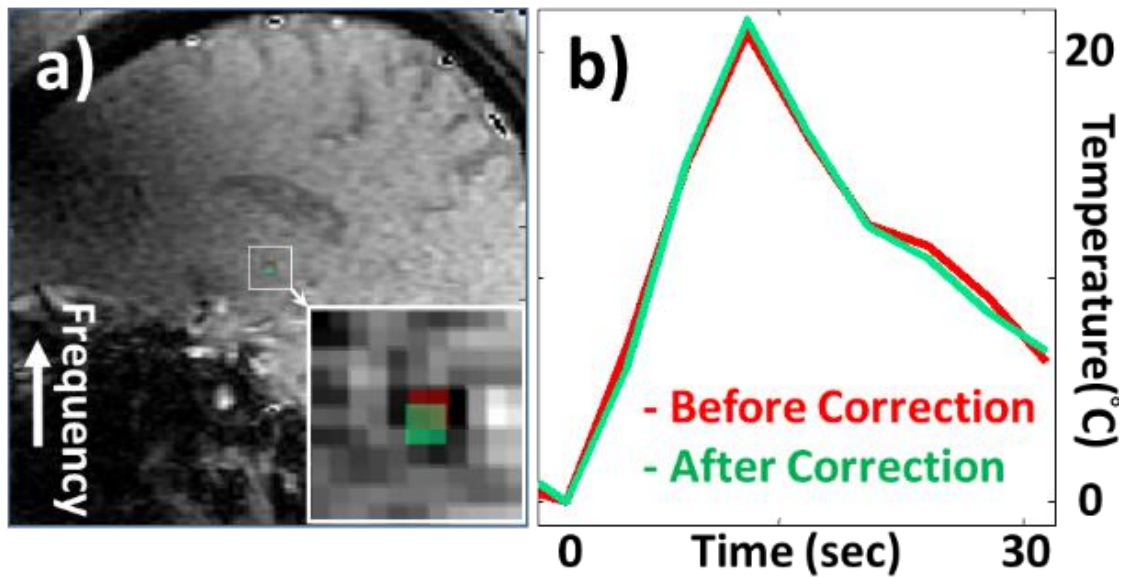

Fig. 10: (a) T maps of focus before and after corrections are overlaid on a brain intensity image, where inset magnifies the shift from its original spot (red) to the corrected spot (green). (b) The temperature changes in time for focus be-fore and after correction.

CONCLUSION

This study highlights the importance of focus correction in MR thermometry for precise targeting in FUS thalamotomy. Our findings demonstrate that both k-space shifts and chemical shift effects contribute significantly to focus location errors. By accounting for both factors, we were able to improve the accuracy of thermal hotspot localization, ensuring a better match between the visualized hotspots and the actual lesion location.

Therefore, we recommend incorporating focus correction, considering both k-space and chemical shift effects, into FUS thalamotomy procedures to enhance targeting precision and minimize the risk of complications. This approach can ultimately contribute to improved patient outcomes and further advance the clinical application of FUS thalamotomy for the treatment of movement disorders.